\documentclass[fleqn,usenatbib]{mnras}

\usepackage{newtxtext,newtxmath}

\usepackage[T1]{fontenc}
\usepackage{ae,aecompl}

\usepackage{graphicx}	
\usepackage{amsmath}	
\usepackage{amssymb}	
\usepackage{gensymb}
\usepackage{enumitem}

\newcommand{\nustar}{\textit{NuSTAR }}

\newcommand{\swift}{{\it Swift }}

\newcommand{\maxi}{MAXI~J1820+070 }
\newcommand{\astrosat}{\textit{AstroSat }}

\title[Broadband X-ray spectrum of \maxi]{A spectral study of the black hole X-ray binary MAXI J1820+070 with \astrosat and \nustar}

\author[S. Chakraborty et al.]{
Sudip Chakraborty$^{1}$\thanks{E-mail: sudip.chakraborty@tifr.res.in (TIFR)},
Nilam Navale$^{2,3}$, 
Ajay Ratheesh$^{4,5,6}$,
Sudip Bhattacharyya$^{1}$ \\
$^{1}$Department of Astronomy and Astrophysics, Tata Institute of Fundamental Research, Mumbai 400005, India\\
$^{2}$University Department of Physics, University of Mumbai, Kalina, Santacruz, Mumbai 400 098, India\\
$^{3}$Ramniranjan Jhunjhunwala College, Ghatkopar, Mumbai 400086, India\\
$^{4}$Department of Physics, University of Rome 'Tor Vergata',
via della Ricerca Scientifica 1, 00133, Rome, Italy.\\
$^{5}$INAF-IAPS, Via del Fosso del Cavaliere 100, I-00133 Rome, Italy\\
$^{6}$Department of Physics, University of Rome 'La Sapienza', P. le A. Moro 2, 00185 Rome, Italy\\
}
\date{Accepted XXX. Received YYY; in original form ZZZ}

\pubyear{2015}

\begin{document}
\label{firstpage}
\pagerange{\pageref{firstpage}--\pageref{lastpage}}
\maketitle

\begin{abstract}
{MAXI J1820+070} is a newly discovered transient black hole X-ray binary, which showed several spectral and temporal features. In this work, we analyse the broadband X-ray spectra from all three simultaneously observing X-ray instruments onboard \textit{AstroSat}, as well as contemporaneous X-ray spectra from \textit{NuSTAR}, observed during the hard state of {MAXI J1820+070} in March 2018. Implementing a combination of multi-colour disc model, relativistic blurred reflection model \textsc{relxilllpCp} and a distant reflection in the form of  \textsc{xillverCp}, we achieve reasonable and consistent fits for \textit{AstroSat} and \textit{NuSTAR} spectra. The best-fit model suggests a low temperature disc ($kT_{\rm in} \sim 0.3$ keV), iron overabundance ($A_{\rm Fe} \sim 4-5$ solar), a short lamp-post corona height ($h \lesssim 8 R_{\rm g}$), and a high corona temperature ($kT_{\rm e} \sim 115-150$ keV). Addition of a second Comptonisation component leads to a significantly better fit, with the $kT_{\rm e}$ of the second Comptonisation component being $\sim 14-18$ keV. Our results from independent observations with two different satellites in a similar source state indicate an inhomogeneous corona, with decreasing temperature attributed to increasing height. Besides, utilising the broader energy coverage of \textit{AstroSat}, we estimate the black hole mass to be $6.7-13.9 \ M_{\odot}$, consistent with independent measurements reported in the literature.
\end{abstract}




\section{Introduction} \label{sec:intro}

Accreting black hole X--ray binaries are unique astrophysical laboratories to probe matter in extreme conditions. Based on the mass of the companion star, they can be classified into High Mass X--ray Binary (HMXB) and Low Mass X--ray Binary (LMXB). In black hole LMXBs, the main source of power is the gravitational energy released by matter accreted from the companion low mass star onto the black hole via Roche lobe overflow (e.g. \citet{FKR_2002}). Based on the long-term temporal evolution of the X-ray emission, LMXBs can be further categorised into persistent and transient sources. Almost all black hole LMXBs are transients \citep{Done_2007}. The transient LMXBs can be in a quiescent state for a long time, probably undetected until they go into outbursts increasing intensities by few orders of magnitude. They exhibit various source states, such as Low Hard State (LHS), High Soft State (HSS) and High Intermediate state (HIS), which can be identified from the Hardness-Intensity Diagram (HID) \citep{Belloni2000_2000A&A...355..271B,RemillardandMcClintock_2006ARA&A..44...49R}. The changes in the spectral states could be attributed to the change in the geometry of the accretion disc \citep{RemillardandMcClintock_2006ARA&A..44...49R}.

In terms of the X-ray spectrum, the HSS is dominated by a multi-colour black-body emission, attributed to the accretion disc. This is interpreted as the disc being very close to the innermost circular stable orbit (ISCO), leading to high disc temperatures \citep{Belloni2000_2000A&A...355..271B,RemillardandMcClintock_2006ARA&A..44...49R}. On the other hand, the LHS is mainly dominated by a variable power-law component, with some hint of low-temperature disc component. The lower disc temperature is interpreted as the disc being truncated at a much larger radius (see, however, \citet{Reis_2010}). The power-law component is generally attributed to a spatially compact region located above the central region of the accretion disc, named ``corona'' \citep{Reynolds_2014}. In this scenario, the hot corona irradiates the accretion disc, producing fluorescent and backscattered radiation. The re-emission from the irradiated disc is thought to be the (thermal) soft excess $\sim 1$ keV, broad iron line complex, and Compton hump around $\sim$30 keV \citep{Fabian_2016}. In LHS, the Comptonising corona and the reflection from the disc give rise to the power-law continuum and reflection features, respectively. While in HSS, the corona weakens or disappears. Thus, LHS is dominated by power-law emission, whereas the HSS is dominated by the higher temperature disc emission. This interpretation of states and state transitions as changes in accretion disc geometry is being challenged in recent years. For example, \cite{ErinKara_2019Natur.565..198K} have proposed the idea of contracting corona as opposed to changing disc truncation radius, as an explanation of states and state transitions. The exact geometry of the corona is still unclear. While a lamp-post geometry is often used for computational simplicity, this may not sufficiently reflect the actual scenario \citep{Chauvin_2018}. Finally, \cite{Yamada_2013}, \cite{Basak_2017} have proposed more general, inhomogeneous structure of the corona.

\maxi is a recently discovered bright X-ray transient, detected with Monitor of All-sky X-ray image, or $MAXI$ \citep{maxi_mission} on board the International Space Station (ISS) on 11th March 2018 at 12:50 UT \citep{maxi_detection_j1820}. Its optical counterpart, named ASSASN-2018ey, was observed with All-Sky Automated Survey for SuperNovae (ASAS-SN) \citep{Tucker_maxij1820,Denisenko_optical_j1820}. A consecutive \swift/BAT trigger led to a follow up monitoring of the source with XRT, which located the X-ray source at RA (J2000) = +18h 20m 21.88s and Dec (J2000) = +07d 11m 08.3s  with an uncertainty of 1.5'' \citep{2018ATel11406....1K}.

\begin{figure}
\centering
    \includegraphics[width=\linewidth]{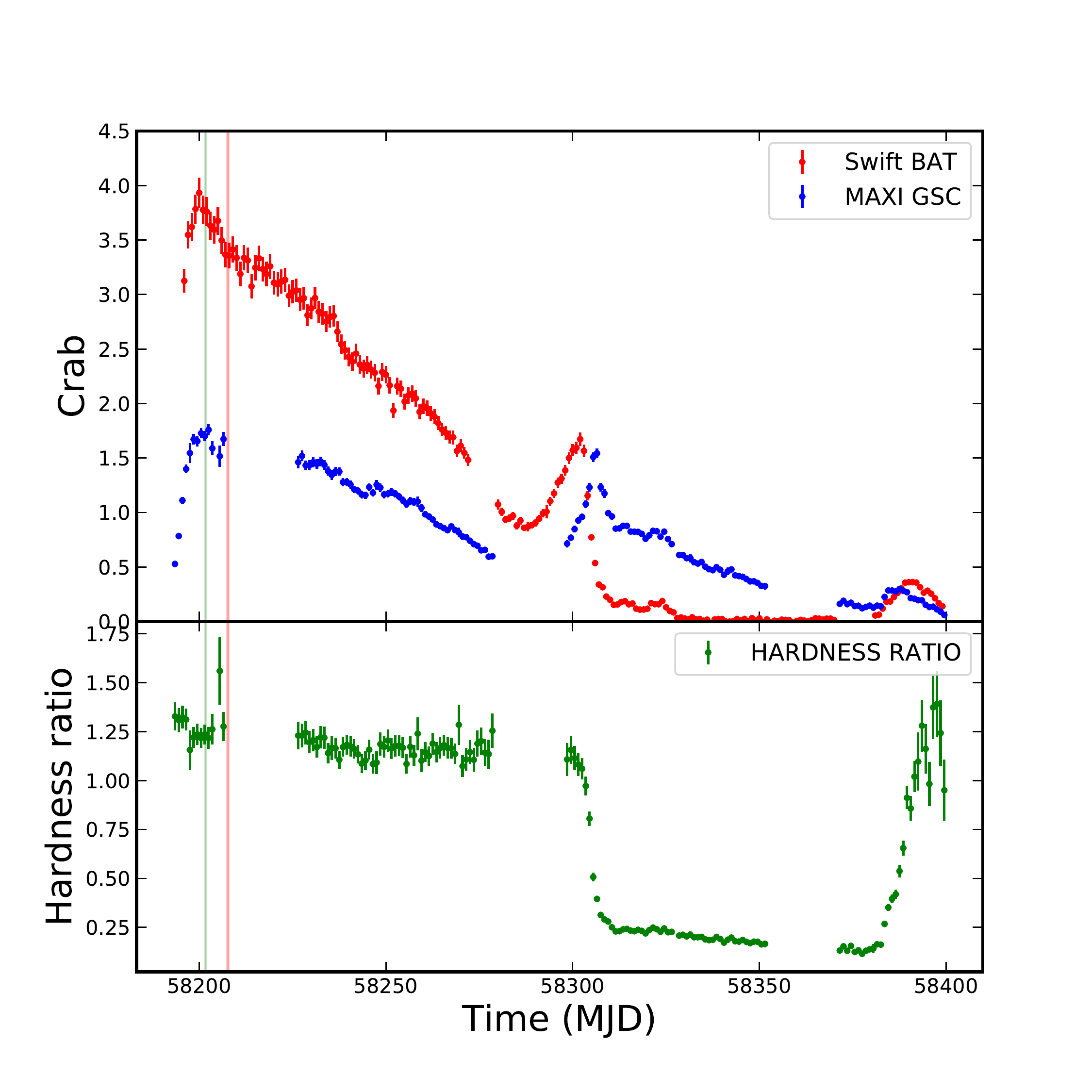}
      \caption{Upper panel : The red and  blue points show the Crab normalised lightcurve of \maxi obtained using \swift/BAT and $MAXI$/GSC in energy range 15-50 keV and 4-10 keV. Lower panel: Crab normalised Hardness ratio (4-10 keV/2-4 keV) obtained using $MAXI$/GSC. The red and green vertical bands mark the epochs observed with \astrosat and \nustar, respectively. }
    \label{bat_maxi_lc}
\end{figure}

During its 2018 outburst, the object was subsequently observed in multiple wavelengths: from radio \citep{2018ATel11420....1B, 2018ATel11439}, to infrared \citep{2018ATel11451....1C,2018ATel11458....1M}, to optical \citep{2018ATel11418....1B, 2018ATel11421....1L,2018ATel11424....1B,2018ATel11425....1G,2018ATel11426....1S,2018ATel11437....1G}. Optical observations \citep{Baglio_2018ATel11418....1B} along with hard X-ray power-law spectrum, and large amplitude in the broadband power spectrum \citep{2018ATel11423....1U} suggested that the source is a Black Hole Low mass X-ray Binary (BH LMXB). After the beginning of the outburst, \maxi underwent rapid increase in flux in both soft and hard X-rays, with hardness ratio remaining more or less constant \citep{2018ATel11427....1D}. Subsequent optical observations revealed an optical period of $\sim$3.4 hr \citep{2018ATel11596....1R}, as well as correlations between X-ray and optical brightness. \citep{PaiceATel_2018ATel11432....1P,paice2019_optical_xraylag, Ateltownsend_2018ATel11574....1T, Yu_atel1_2018ATel11510....1Y}. Further sub-millimetre detection \citep{2018ATel11440....1T} and flat radio spectrum \citep{2018ATel11440....1T} hinted the launch of a relativistic jet. 
After the initial rise in intensity, \maxi underwent a slow decay from around MJD 58200. 
The hard X-ray flux dropped sharply around MJD 53805, causing a state transition to the soft state. The source remained in the soft state for around 60 days before transiting back to the hard state around MJD 58390, before fading away into quiescence after MJD 58400 (Fig. \ref{bat_maxi_lc}).

During the LHS, \maxi had a hard spectrum with a photon index of $\sim$1.5 \citep{2018ATel11427}. Low-frequency quasi-periodic oscillations (QPOs) were detected in Optical and X-ray wavelengths (10-50 mHz) \citep{gandhi_2018ATel11437....1G,Mereminskiy_2018ATel11488....1M,Yu_2018ATel11510....1Y}. Strong optical and X-ray short-term variabilities on time scales of less than 1 s were also reported \citep{gandhi_2018ATel11437....1G, Sako_2018ATel11426....1S}.
\cite{ErinKara_2019Natur.565..198K} performed spectro-temporal study of \maxi and detected reverberation time lags between 0.1--1.0 keV and 1.0--10.0 keV energy bands. The authors have observed a corona height <5$R_{\rm g}$ and an inner disc radius <2$R_{\rm g}$ (where $R_{\rm g}$ is the gravitational radius of the black hole, defined as $R_{\rm g}=GM/c^2$). 
They also claimed that with the evolution of the outburst the corona became compact and shifted close to the compact object. Spectral analysis during the LHS by \cite{Buisson_2019MNRAS.490.1350B} not only supported this claim but also suggested a correlation between X-ray luminosity and coronal temperature. However, a general relativistic simulation using the spectral analysis indicated that the spectral softening was due to the change in the inner disc radius rather than the coronal geometry \citep{Roh_2020AAS...23537902R}. The mass function of the binary system was reported to be 5.18 $\pm$ 0.15 \(M_\odot\) \citep{Torres_massEstimate_2019ApJ...882L..21T}. By constraining the inclination angle to be $69^{\circ}$ < $i$ < $77^{\circ}$, the black hole mass was estimated to be 7-8 \(M_\odot\) \citep{Torres_massEstimate_2019ApJ...882L..21T}. Using radio parallax method, \cite{Atri2020} estimated the accurate distance to the source to be 2.96$\pm$0.33 kpc, further constraining the black hole mass of the source to 9.2 $\pm$ 1.3 \(M_\odot\). Analysing the HSS spectra, \citet{Fabian_2020} found the black hole mass to be $\sim 7-8M_\odot$, and the black hole spin to be moderately low (dimensionless spin parameter, $a$, between -0.5 to +0.5). Using DR2  data of Gaia, \cite{Gandhi_2019} found the source distance to be  $3.46^{+2.18}_{\rm −1.03}$ kpc.  By  analysing  the  soft  X-ray  data  from NICER during the rising phase of outburst, \cite{2018ATel11423....1U} found  the  Galactic  extinction  ($N_{\rm H}$) to  be $1.5\times10^{21} \rm cm^{-2}$.

In this work, we investigate the broadband spectral characteristics of the hard state spectra of \maxi during its 2018 outburst. We use the data from \astrosat observation during March 30-31, 2018 and a suitably chosen \nustar observation during March 24-25, 2018. The \astrosat spectra (without the Cadmium Zinc Telluride Imager (CZTI)) were presented with a simple model fit (and with greater systematic uncertainty assumed) in \citet{Mudambi_2020}. 
Of the broadband X-ray satellites currently available, \nustar and \astrosat provide good energy resolution in hard X-ray range ($>10$keV), with \astrosat covering a broader energy range. Here we utilise this opportunity, to systematically and uniformly study the broadband spectra of this BHXB from these independent satellite instruments using detailed spectral models for the first time.
The paper is organised as follows. In section \ref{sec:obs} we describe the data reduction procedure of \nustar, as well as all three pointing X-ray instruments on board \astrosat, namely Soft X-ray Telescope (SXT), The Large Area X-ray Proportional Counter (LAXPC) and CZTI. In section \ref{sec:lc}), we describe the $MAXI$/GSC and \swift/BAT lightcurves and HID generation. We then describe the states and state transition, as well as the \nustar data selection based on the mentioned HID in section \ref{sec:hid}. Furthermore, we present an in-depth spectral analysis of \nustar and \astrosat data in section \ref{sec:wo refl} and \ref{sec:w refl}, followed by an estimation of the black hole mass based on the same spectral fit in section \ref{sec:mass}. Finally, we summarise our results and discuss their implications in section \ref{sec:discussion}.

\begin{table}
\centering
\begin{tabular}{ |c|c|c|c| } 
\hline
Instrument & Obs ID & Obs. date  & Exposure (s) \\
        & (yyyy-mm-dd)   &  \\
\hline
\nustar & 90401309010 & 2018-03-24/25 & 2660 \\
\astrosat & 9000001994 & 2018-03-30/31 & 11768 \\
\hline
\end{tabular}
\caption{\nustar and \astrosat observation details. In case of \astrosat, we have mentioned only the SXT exposure time. The LAXPC and CZTI exposure times are higher by more than a factor of 2.}
\label{}
\end{table}

\section{Observation and data reduction} \label{sec:obs}

\astrosat, the first Indian dedicated astronomy satellite, was successfully launched on 2015 September 28, carrying five scientific instruments on board \citep{Singh_ASTROSAT_2014}: the focusing Soft X-ray Telescope (SXT), Large Area X-ray Proportional Counters (LAXPCs, 3 units), the hard X-ray Cadmium Zinc Telluride Imager (CZTI), the all-sky monitor called Scanning Sky Monitor (SSM) and the Ultraviolet Imaging Telescopes (UVIT). 
\astrosat observed \maxi in a hard state during the outburst on 2018 March 30-31, and we use the data from all three X-ray instruments (SXT, LAXPC and CZTI) in the present work.

\subsection{SXT}
\label{sec:sxt} 

SXT (\cite{Singh_2016,Singh_2017}) is a grazing incidence X-ray telescope on board \astrosat, with a focal length of 2 m. 
It covers the energy range of 0.3-8 keV and has a field of view of $\sim 40^{'}$ \citep{Singh_ASTROSAT_2014}.

The SXT data are acquired in the PC mode and is significantly piled up. To minimise the effect of the pile-up, source spectrum is extracted from an annulus between 6' and 15' from the centre of the image (the details of the region selection for avoiding pile-up, are described in Appendix ~\ref{appendix:pile-up}). The deep blank sky background spectrum, provided by the instrument team\footnote{\url{https://www.tifr.res.in/~astrosat_sxt/index.html}}, is used for the spectral modelling. 

\subsection{LAXPC}
\label{sec:laxpc} 
LAXPC \citep{2016ApJ...833...27Y} is a proportional counter array on board \astrosat{} \citep{2006AdSpR..38.2989A, Singh_ASTROSAT_2014} with a large effective area, which observes sources in Event Analysis (EA) mode with absolute time resolution of 10 $\mu$s in the energy range
of 3.0-80.0 keV. The extraction of lightcurve and spectrum from level 1 data is done by laxpcsoftv3.1 \footnote{\url{http://www.tifr.res.in/~astrosat_laxpc/LaxpcSoft_v1.0/antia/laxpcsoftv3.1_04Sept2019.tar.gz}} along with background lightcurve and background spectrum. Among the three units of LAXPC, we have considered only one unit. As LAXPC30 has gain instability issue caused by a gas leakage and LAXPC10 was showing unpredictable HV variations, the results reported in this paper are from LAXPC20. We have also extracted lightcurves in different energy bands corresponding to \nustar, \swift/BAT and $MAXI$/GSC energies to plot HID as shown in Fig. \ref{fig:2}.   

\subsection{CZTI}
\label{sec:czti} 
The level 1 data of \astrosat CZTI is reduced to level 2 cleaned events, spectra, and lightcurve using the tool \textit{cztipipeline} of cztipipeline Ver 2.1\footnote{\url{http://astrosat-ssc.iucaa.in/?q=data\_and\_analysis}}. The total light curve before background reduction is then checked for any instrumental anomalies near the South Atlantic Anomaly (SAA). The time intervals very close to the SAA, where a sharp increase in the count-rate is observed, are excluded. Only Quadrant 0 in CZTI is used for the spectrum, as the instrumental response features in the other quadrants are yet to be understood. Further, module $13$ in Quadrant zero is excluded as with Lower Limit of Detection (LLD) at around $50$ keV, it does not span the entire spectral region of interest from the CZTI in this particular case.

\subsection{\nustar}
\label{sec:nustar} 
\maxi was observed with \nustar \citep{Harrison_NuSTAR_2013} during 2018 March 24-25 (the reasons of this data selection are detailed in section \ref{sec:state} and section \ref{sec:discussion}). The \nustar data are processed using  v.1.8.0 of the NuSTARDAS pipeline with \nustar CALDB v220171002. After filtering background flares due to enhanced solar activity by setting saacalc = 2, saamode = OPTIMIZED, and tentacle = no in NUPIPELINE, the effective exposure times are 2.6 and 2.8 ks for the two focal plane modules FPMA and FPMB, respectively. The source spectra are extracted from a circular region of the radius 180$^{''}$ centred on the source location. The background spectra are extracted from a blank region on the detector furthest from the source location to avoid source photons.
The spectra are grouped in $ISIS$ \citep{Houck_2000} version 1.6.2-41 to have a signal-to-noise ratio of at least 50 per bin, similar to \cite{Buisson_2019MNRAS.490.1350B}.

\subsection{$MAXI$/GSC and \swift/BAT Lightcurve} \label{sec:lc}
The Burst Alert Telescope on board Neil Gehrels \swift Observatory (\swift/BAT) and Gas Slit Camera on board Monitor of All-sky X-ray Image ($MAXI$/GSC) continuously traced the source throughout the outburst phase. The daily averaged lightcurves from \swift/BAT in 15 to 50 keV have been obtained from \url{https://swift.gsfc.nasa.gov/results/transients/} and is normalised by the Crab count-rate in \swift/BAT. A very sharp increase in flux is seen during the initial phase with a gradual decrease. Another peak is also seen around MJD 58300. For $MAXI$/GSC, \url{http://maxi.riken.jp} provided the lightcurves in 2-4 keV and 4-10 keV. The lightcurves are further normalised in respective bands by the Crab count-rate in $MAXI$/GSC. Further the hardness ratios are also obtained using the Crab normalised count-rates, the hard band being 4-10 keV and soft being 2-4 keV. A sharp rise is seen also in the $MAXI$/GSC (4-10 keV) during the initial phase but the hardness ratio remains constant. A sharp decrease is also observed around MJD 58300, and a further sharp rise is observed around  MJD 58380. \astrosat and \nustar made pointed observations during the outburst marked by red and green vertical bands.

\subsection{Hardness Intensity Diagram Generation from $MAXI$} \label{sec:hid}
\begin{figure}
\centering
    \includegraphics[width=\linewidth]{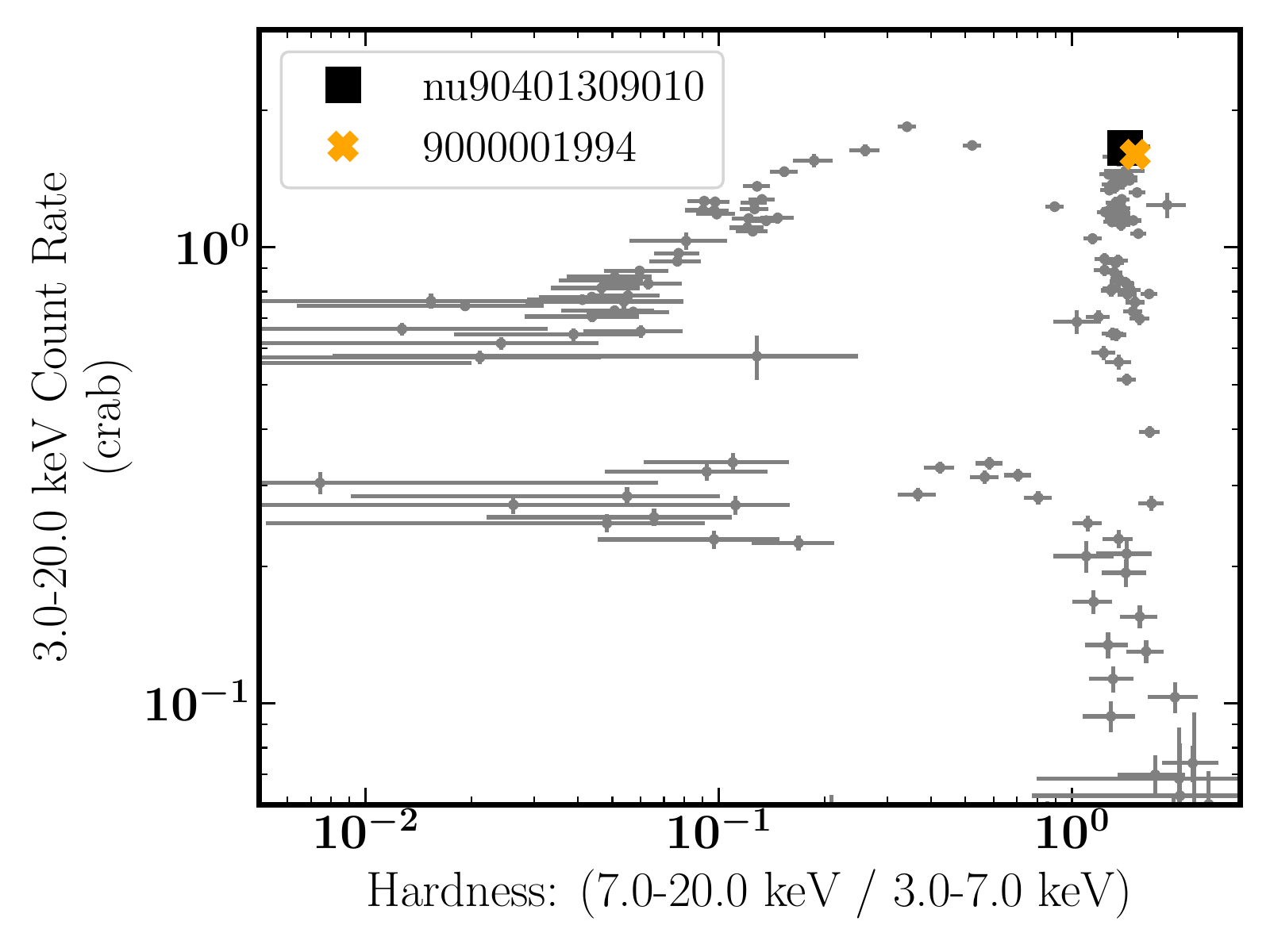}
      \caption{Hardness intensity diagram (HID) of \maxi with 1 day averaged $MAXI$/GSC monitoring observations during its first outburst. The scattered grey circles indicate the overall evolution of the source from MJD 58188 to MJD 58734. The \astrosat observation of interest is marked with an orange cross on the HID and the nearest \nustar observation which have consistent spectral properties is marked with a black square. The size of the error bars is similar to that of marker size for these two observations. See section ~\ref{sec:state} for a discussion on the HID.}
    \label{fig:2}
\end{figure}

The hardness intensity diagram (HID) for \maxi is been generated using daily monitoring data from $MAXI$/GSC (Fig. \ref{fig:2}). For hardness ratio (Defined as the background-subtracted count-rate ratio in quoted energy bands), we choose 3.0-7.0 keV and 7.0-20.0 keV energy bands and consider intensity from 3.0 keV to 20.0 keV. Each point in the HID is averaged over one day. The pointed observations of the \astrosat{} (orange cross) and the \nustar{} (black square) reported in this paper are marked over the HID. The data from all three instruments are normalised by Crab counts, for a better comparison. 

\section{Results} \label{sec:results}

\subsection{Spectral State and Transition} \label{sec:state}
Most of the black hole X-ray binaries (BHXBs) are of transient nature and are discovered when outbursts occur. The evolution of an outburst in a BHXB is well depicted by a Hardness Intensity Diagram (HID) \citep{2001ApJS..132..377H, 2005Ap&SS.300..107H, 2006ARA&A..44...49R,2009MNRAS.396.1370F}, which traces a `q-shaped' trajectory moving in the counter clockwise direction. The best-known example is GX 339-4, which undergoes recurrent outbursts and has always followed a `q' shaped path on HID \citep{2005A&A...440..207B,2009MNRAS.396.1370F,2011MNRAS.418.2292M}. The HID is a model-independent tool to study the state transitions in such BHXBs.

We generate the HID for \maxi{} (Fig. \ref{fig:2}), using a daily monitoring $MAXI$/GSC data since the time of its discovery. The source has followed a `q'-shaped track similar to GX 339-4. Various states on HID, as described in section \ref{sec:intro}, are evident in Fig.\ref{fig:2}. We plot the \astrosat observation of our interest over the HID (orange cross), which is found to be in the hard state. During this state both hardness and intensity are high. We also considered one of the nearest \nustar observations (black square). The \nustar observation is also in a similar source hardness.  

\subsection{Spectral Analysis} \label{sec:spectra}

The RMF files for SXT and LAXPC are obtained from the respective Payload Operation Center (POC) websites\footnote{\url{https://www.tifr.res.in/~astrosat_sxt/index.html}}\footnote{\url{https://www.tifr.res.in/~astrosat_laxpc/}}. For SXT, sxt\_pc\_mat\_g0to12.rmf is used as the RMF and sxt\_pc\_excl00\_v04\_20190608.arf is used as the ARF, whereas for LAXPC20 the response file lx20cshm03v1.0.rmf is used. For joint fitting, both the SXT and LAXPC spectral files are binned in GRPPHA to have at least 30 counts per bin to facilitate $\chi^2$ fitting. A 2\% systematic uncertainty, as prescribed by the instrument teams \footnote{\url{https://www.tifr.res.in/~astrosat_sxt/dataana_up/readme_sxt_arf_data_analysis.txt}}, is used for the spectral fitting. For CZTI, the spectral and response files are generated using \textit{cztbindata} and \textit{cztrspgen} of the \textit{cztipipeline}\footnote{\url{http://astrosat-ssc.iucaa.in/uploads/czti/CZTI_level2_software_userguide_V2.1.pdf}} respectively.

The spectral fitting and statistical analysis are carried out using the XSPEC version v-12.9.0n \citep{Arnaud_1996}. Energy ranges of 1.3-7.0 keV and 5-60 keV are used for SXT and LAXPC20 respectively. The photons below 1.3 keV and above 7.0 keV for SXT, and below 5.0 keV and above 60 keV for LAXPC, are ignored to avoid larger systematic errors. For CZTI, the energy range of 30 keV to 120 keV is used. Additionally, a gain correction is applied to the SXT data using the XSPEC command `gain fit' with a slope of unity. The best fit offset value is found to be 34 eV and is used throughout the paper. For the joint fitting between different \astrosat instruments, a cross-normalisation constant (implemented using \textsc{Constant} model in XSPEC) is allowed to vary freely for LAXPC and CZTI and is assumed to be unity for SXT. Similar cross-normalisation is considered between FPMA and FPMB for \nustar  data fit. For \nustar, an energy range between 3 keV and 78 keV is considered for spectral fitting. To avoid the sharp instrumental features (as reported by \cite{Xu_J1535_2018}), energies between 11-12 and 23-28 keV are excluded (\cite{Buisson_2019MNRAS.490.1350B}). All the models, as described below, include the Galactic absorption through the implementation of the \textsc{TBabs} model. The corresponding abundances are set in accordance with the \cite{Wilms_2000} photoelectric cross-sections. The neutral hydrogen column density ($N_{\rm H}$) is fixed to $1.5 \times 10^{21} \ \rm cm^{-2}$ \citep{2018ATel11423....1U}  for all the described models. All parameter uncertainties are reported at the 90\% confidence level for one parameter of interest.

\subsubsection{Spectral fits without reflection} \label{sec:wo refl}

\begin{figure}
\centering
	\includegraphics[width=1.0\linewidth]{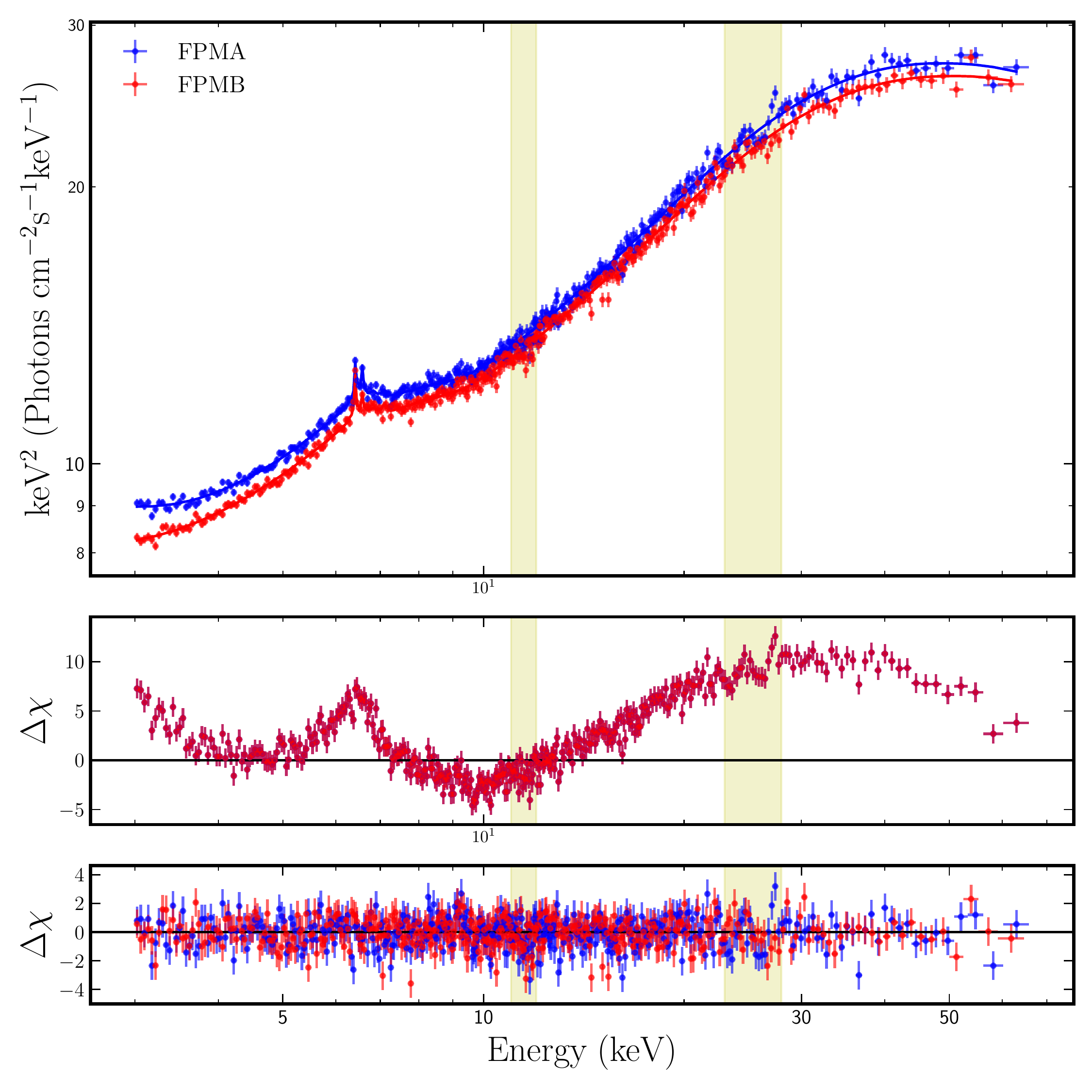}
  	\caption{Results of 3-78 keV \nustar spectral fit of \maxi (see sections~\ref{sec:w refl},~\ref{sec:discussion} for further details). Top panel: the unfolded \nustar spectra. The blue and red points indicate FPMA and FPMB data, respectively. The spectra are folded with M2 model here. The yellow vertical bands denote the energy ranges ignored for spectral fitting due to instrumental features. Middle panel: The residuals of \nustar data, fitted with a fiducial \textsc{TBabs$\times$cutoffpl} model. A soft excess, broad iron line and a Compton hump around $\sim$30 keV are visible. Bottom panel: Residuals from M2 model fit.}
    \label{fig:nustar}
\end{figure}

First of all, to demonstrate the reflection features, we fit the \nustar spectra with an absorbed cutoff power-law model, \textsc{TBabs$\times$cutoffpl} in XSPEC notation. For this, we only consider the energy intervals of 3-4, 8-12, and 40-78 keV, where reflection from the disc has minimal effect. As displayed in the second panel of Fig. \ref{fig:nustar}, a broad iron Fe K-$\alpha$ emission with a narrow core, as well as a Compton hump peaking around 30 keV, are evident in the residuals. The slight difference ($\sim$1\%) between FPMA and FPMB below 5 keV, is within the calibration accuracy of \nustar \citep{Madsen_2015}.

For a better explanation of the observed broadband energy spectra, we fit the \nustar data with a model comprising of a combination of multicoloured disc
black-body (diskbb: \cite{Mitsuda_1984},\cite{Makishima_1986}) and thermal Comptonisation (Nthcomp: \cite{Zdziarski_1996},\cite{Zycki_1999}), \textsc{TBabs$\times$(diskbb + Nthcomp)} in XSPEC notations. The seed photon temperature in \textsc{Nthcomp} is set to the innermost temperature ($T_{\rm in}$) of the \textsc{diskbb} component. In the energy range mentioned before, we get the best-fit \textsc{diskbb} $T_{\rm in}$ to be $0.77^{0.06}_{\rm -0.05}$ keV, a power-law index ($\Gamma$) of $1.58 \pm 0.01$, and the electron temperature $kT_{\rm e}$ to be $19.2 \pm 0.3$ keV. The fit results in a reduced $\chi^2$/dof of 404.2/211. The value of the cross-normalisation factor between FPMA and FPMB is found to be $0.96 \pm 0.01$ ($\sim 4\%$), which is within the accepted limit of $\le 5\%$ (\cite{Madsen_2015},\cite{Marcotulli_2017}). Due to the limited spectral coverage assumed here, the value of $kT_{\rm e}$ is somewhat low. To get a better fit and to explore the possibility of additional spectral components, we add another \textsc{Nthcomp} model to the existing model. This double Comptonisation model gives a much better fit, with a $\chi^2$/dof of 227.8/208. The fit results in a segregation of corona temperatures, with the best-fit $kT_{\rm e}$ being $27.4^{2.7}_{\rm -2.3}$ keV and $1.35^{0.02}_{\rm -0.05}$ keV. The $T_{\rm in}$ of the disc is found to be $0.42 \pm 0.01$ keV. Note that this exercise is for a  demonstration purpose only and the derived values are unreliable/unphysical as these fits within narrow energy bands do not include the effect of reflection, which we will discuss in greater depth in section \ref{sec:w refl}. In fact, the second Comptonisation component in this model could be fitting for the Compton hump of the reflection (though we have tried to avoid this by ignoring the corresponding energy range) rather than a true second Comptonisation. This shows the need for proper reflection modellings. Nevertheless, the important qualitative indication from this exercise is the possible existence of a second Comptonisation component, with a lower corona temperature.

Inspired by this result, we investigate the possibility of a double Comptonisation scenario in greater detail in section \ref{sec:w refl}. Until now, we have not explored the reflection. In the following section, we perform an in-depth reflection modelling of both the \nustar and \astrosat broadband spectra.

\subsubsection{Spectral fits with reflection} \label{sec:w refl}

For a detailed investigation of the broadband spectra including the reflection features, we use the self-consistent relativistic disc reflection models from \textsc{relxill} model suite (relxill v1.2.0 : \cite{Dauser_2014}, \cite{Garcia_2014}). To minimise the number of free parameters, we assume a lamp-post geometry of the Comptonising corona and use the model \textsc{relxilllpCp} which internally includes an \textsc{Nthcomp} continuum. However, we only include the reflected flux from the \textsc{relxilllpCp} component (by setting the model parameter refl\_frac$<0$). We add an explicit \textsc{Nthcomp} component to represent the continuum. The seed photon temperature of the \textsc{Nthcomp} component is tied to the inner accretion disc temperature (\textsc{diskbb} $T_{\rm in}$), and the electron temperatures ($kT_{\rm e}$) and spectral indices ($\Gamma$) of both the \textsc{Nthcomp} and \textsc{relxilllpCp} components are tied together. As before, the  value of $N_H$ is fixed to $1.5 \times 10^{21} \ \rm cm^{-2}$, and the $T_{\rm in}$ from the additional \textsc{diskbb} component is used as the seed temperature for the internal \textsc{Nthcomp} continuum. 

The inner radius ($R_{\rm in}$) of the thin accretion disc, and the dimensionless black hole spin parameter ($a$), which are reflection spectrum parameters, are degenerate. For simplicity in spectral fitting, we assume a maximally spinning black hole ($a = 0.998$), and fit for $R_{\rm in}$ (for further discussion, refer to section~\ref{sec:discussion}). We fix the outer edge of the accretion disc ($R_{\rm out}$) at $400R_{\rm g}$ (where $R_{\rm g}$ is the gravitational radius of the black hole, defined as $R_{\rm g}=GM/c^2$). On the other hand, we keep the inclination angle free (the data has sufficient signal for this purpose, as shown by \citet{Buisson_2019MNRAS.490.1350B}). The reflection fraction can thus be self-consistently determined by fitting $R_{\rm in}$ and the height of the lamp-post ($h$) from ray-tracing calculations \citep{Dauser_2014}. To account for the narrow core of the Fe-K$\alpha$ line, we use the unblurred reflection model \textsc{xillverCp} \citep{Garcia_2010}. Similar to our implementation of \textsc{relxilllpCp}, we use \textsc{xillverCp} only as a reflection component. The refl\_frac of \textsc{xillverCp} is fixed at -1, as only insignificant variations are found in the subsequent fits if the refl\_frac is allowed to vary freely.
We tie the $\Gamma$ and $kT_{\rm e}$ of the input continuum, as well as the iron abundances ($A_{\rm Fe}$) and inclination angles ($\theta$) of the \textsc{relxilllpCp} and the \textsc{xillverCp} components. The ionisation parameters ($\xi=L/nR^2$, where $L$ is the ionising continuum luminosity, $R$ is the distance to the ionising source and $n$ is the gas density) are allowed to be free for both the components. For \textsc{xillverCp}, we allow the $\rm{log} \xi$ to be non-zero, thus allowing for distant reprocessing by ionised gas. Thus, the total model setup is \verb'TBabs'$\times$\verb'(diskbb+Nthcomp(1)+relxilllpCp(1)+xillverCp)'. This becomes our model M1 in table \ref{tab:param}. 

Inspired by the two Comptonisation scenario presented in section \ref{sec:w refl}, we further add an additional \textsc{Nthcomp} to the M1 model, reflected through the \textsc{xillverCp} component. For this purpose, we link the $\Gamma$ and $kT_{\rm e}$ between the \textsc{xillverCp} and the new \textsc{Nthcomp} components. The seed temperature of the new \textsc{Nthcomp} is again linked to the $T_{\rm in}$ of the \textsc{diskbb}. Thus, the resulting model \verb'TBabs'$\times$\verb'(diskbb+Nthcomp(1)+relxilllpCp(1)+xillverCp+' \\
\verb'Nthcomp(2))' becomes our model M2 in table \ref{tab:param}.

To summarise, the two main models (in table \ref{tab:param}) considered in our works are as follows:
\begin{enumerate}[leftmargin=\parindent,align=left,labelwidth=\parindent,labelsep=0pt]
	\item \verb'TBabs'$\times$\verb'(diskbb+Nthcomp(1)+relxilllpCp(1)+xillverCp)': This forms the M1 model in table \ref{tab:param}. Here, the \textsc{Nthcomp} acts as the continuum, and the \textsc{relxilllpCp} and \textsc{xillverCp} act solely as reflection, both reflecting the \textsc{Nthcomp} component. The seed temperature of the \textsc{Nthcomp} is tied to the \textsc{diskbb} $T_{\rm in}$, the $kT_{\rm e}$ and $\Gamma$ of the \textsc{Nthcomp}, \textsc{relxilllpCp} and \textsc{xillverCp} are tied to each other. The $A_{\rm Fe}$ and $\theta$, of the \textsc{xillverCp} are tied to that of \textsc{relxilllpCp} component, while its $\xi$ is allowed to vary freely. Basically, this model comprises of a single Comptonisation continuum being reflected from the disc by two different reflection components, one blurred and the other unblurred.
	\vspace{0.4 cm}
	\item \verb'TBabs'$\times$\verb'(diskbb+Nthcomp(1)+relxilllpCp(1)+xillverCp+' \\
          \verb'Nthcomp(2))':  This makes the M2 model in table \ref{tab:param}. Here, the \textsc{Nthcomp(1)} acts as the hard/primary continuum, and the \textsc{relxilllpCp} reflects the \textsc{Nthcomp(1)} component. The seed temperature of the \textsc{Nthcomp(1)} is tied to the \textsc{diskbb} $T_{\rm in}$, the $kT_{\rm e}$ and $\Gamma$ of the \textsc{Nthcomp} and \textsc{relxilllpCp} are tied to each other. The \textsc{Nthcomp(2)} component denotes the second, soft Comptonisation component; reflected by the \textsc{xillverCp} component. The seed temperature of the \textsc{Nthcomp(2)} is tied to the \textsc{diskbb} $T_{\rm in}$, and the $kT_{\rm e}$ and $\Gamma$ of the \textsc{xillverCp} are tied to that of the  \textsc{Nthcomp(2)} component. Just like in M1, the $A_{\rm Fe}$ and $\theta$, of the \textsc{xillverCp} are tied to that of \textsc{relxilllpCp} component, while its $\xi$ is allowed to vary freely. More crucially, the $kT_{\rm e}$ and $\Gamma$ of the two Comptonisation components are not tied to each other. Essentially, this model comprises of two Comptonisation continuum being reflected from the disc by two different reflection components, one blurred and the other unblurred, having different temperatures (and potentially two different lamp-post heights). This better mimics a realistic corona, allowing for a temperature structure.
\end{enumerate}

To make a reasonable comparison with the \nustar spectral fit by \cite{Buisson_2019MNRAS.490.1350B}, we allow the \textsc{diskbb} $T_{\rm in}$ and normalisation, as well as the spectral index and normalisation of the intrinsic continuum to vary freely between FPMA and FPMB. This takes care of the slight calibration differences between the two modules. We find that apart from the \textsc{diskbb} normalisation, the typical difference between the two modules in the other quantities are less than the stated calibration level \citep{Madsen_2015}. The introduction of the reflection models vastly improves the goodness of fit. For model M1, the $\chi^2$/dof is found to be 715.1/670. The innermost accretion disc temperature is observed to be $0.44^{+0.08}_{\rm -0.07}$ keV, which is lower than the values reported by \cite{Buisson_2019MNRAS.490.1350B}. However, we have to keep in mind that \nustar extends only up to 3.0 keV in lower energy side, and hence is not extremely suitable for accurate estimation of the accretion disc parameters on its own. Most of the other parameters, viz. $\Gamma, R_{\rm in}, \log\xi, A_{\rm Fe}$ are found to be consistent with the parameters in \textsc{relxilllpCp(1)} model fit of epoch 3 spectra in \cite{Buisson_2019MNRAS.490.1350B}. The values of $\Gamma$ are consistent with hard state spectra of black hole X-ray binaries \citep{RemillardandMcClintock_2006ARA&A..44...49R}. Inner disc radius ($R_{\rm in}$, of $6.9^{+0.9}_{\rm -1.0} R_{\rm g}$) extending very close to the Innermost Stable Circular Orbit (ISCO) and super-solar abundances ($A_{\rm Fe}$ of $4.4^{+1.3}_{\rm -0.5}$ time solar) are also observed. The proximity of the inner radius of the disc to the ISCO is consistent with \citet{ErinKara_2019Natur.565..198K,Buisson_2019MNRAS.490.1350B}.
The non-zero ionisation parameter ($\log \xi  =  2.5^{+0.2}_{\rm -0.1} \ \rm \log[erg.cm/s]$) of the  \textsc{xillverCp} indicates the presence of an ionised distant reflection component. However, the lamp-post height is found to be pegged near the lowermost allowable limit. The inclination angle is found to be $29^{+2}_{-8}$ degrees. Additionally, we observe a higher corona temperature ($118^{+29}_{-19}$ keV) than the previously reported values. The best-fit parameter values are presented in the first column of table ~\ref{tab:param}. 

\begin{figure*}
\centering
	\includegraphics[angle=0.0,width=0.49\textwidth]{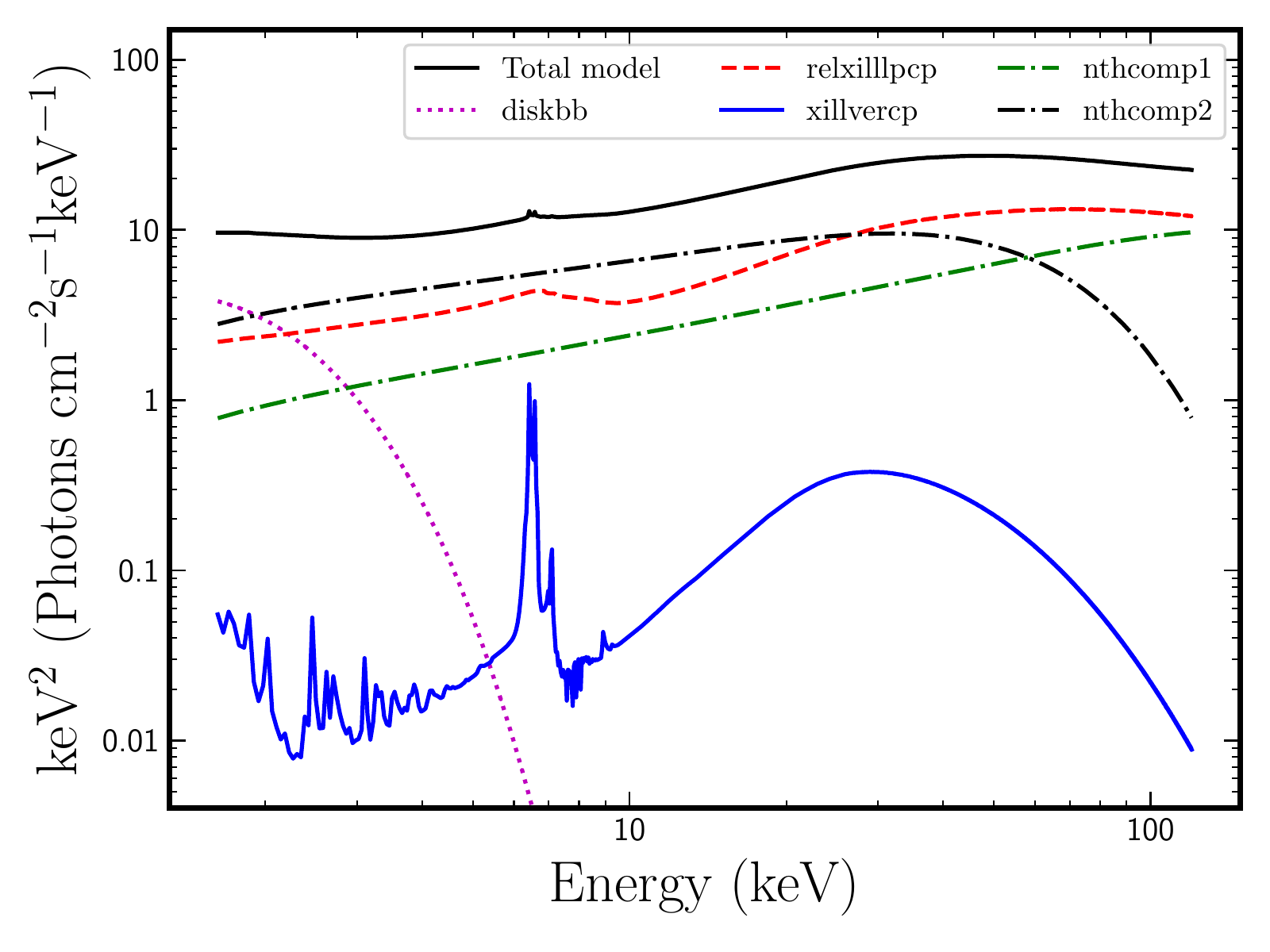}\quad
	\includegraphics[angle=0.0,width=0.49\textwidth]{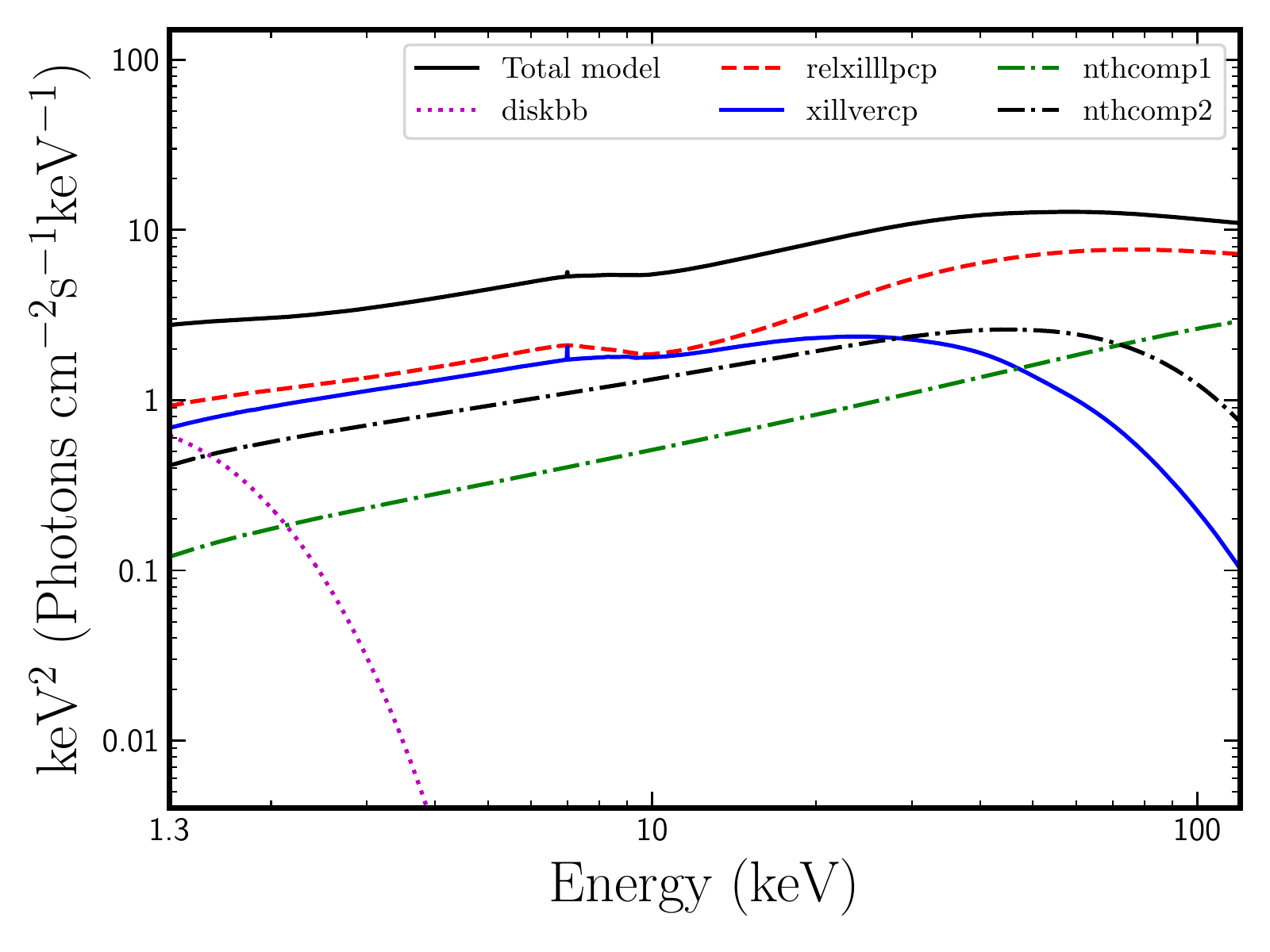}
  	\caption{Best-fit M2 model for \nustar (left panel) and \astrosat (right panel). The dotted magenta curve corresponds to the multi-colour disc component. The dashed red curves and solid blue curves mark the \textsc{relxilllpCp} and \textsc{xillverCp} components, respectively. The intrinsic \textsc{Nthcomp} continua are represented by the green dash-dotted curves. The black dash-dotted curves indicate the second \textsc{Nthcomp} Comptonisation components, with lower rollover energy. 
  	 The spectral components are detailed in section~\ref{sec:w refl}, and the implications discussed in section~\ref{sec:discussion}. }
    \label{fig:eemo}
\end{figure*}

Addition of the second Comptonisation component improves the best-fit significantly. If the $kT_{\rm e}$ of the additional \textsc{Nthcomp} is tied to that of the original \textsc{relxilllpCp} component, the fit results in a $\chi^2$/dof of 688.8/668. The common temperature of the corona is found to be $38^{+3}_{\rm -2}$ keV. This value is similar to that obtained by \citet{Buisson_2019MNRAS.490.1350B}, the implication of which is discussed later in this section and section ~\ref{sec:discussion}. Now, if we untie the two $kT_{\rm e}$ components (thus forming model M2), the $\chi^2$/dof becomes 666.1/667. The addition of the 3 free parameters change the best-fit values. All the disc reflection parameters, including the lamp-post height ($7.41^{+1.94}_{\rm -1.93}R_{\rm g}$), are found to be more or less consistent with the epoch 3 results from \cite{Buisson_2019MNRAS.490.1350B}. The additional Comptonisation component is found to be at a much lower $kT_{\rm e}$ value of $14.0^{+2.1}_{\rm -1.7}$ keV, as compared to $115^{+38}_{\rm -29}$ keV of the primary Comptonisation component.
The best-fit parameter values can be found in the second column in table ~\ref{tab:param}. This segregation of the temperatures of the two coronal components hints towards the existence of an inhomogeneous corona. When the temperatures of the two components are tied, we get an average temperature of $\sim 38$ keV, which \citet{Buisson_2019MNRAS.490.1350B} found. An $F-test$ between the fits with M1 and M2 model yields a very low false-positive probability of $2.9 \times 10^{-10}$, further solidifying the importance of the second Comptonisation component for obtaining a good fit. A detailed representation of the different spectral components can be found in Fig. \ref{fig:eemo} and the \nustar residual with M2 model fit is presented in the bottom panel in Fig. \ref{fig:nustar}.

Similar to the \nustar spectral fits, we model the joint SXT+LAXPC+CZTI spectra. The broader energy range of 1.3-120.0 keV gives us a better handle over the disc parameters and high energy rollover, although the poorer spectral resolution around the iron lines gives a less reliable measurement of the reflection parameters. The best-fit parameters from M1 model fit are found to a little different than the contemporaneous \nustar fit. The inner disc temperature ($0.33^{+0.01}_{\rm -0.01}$ keV) is found to be somewhat higher than the value reported in \cite{Mudambi_2020}, and close to the is typical $T_{\rm in}$ values for the hard spectral state in black hole X-ray binaries (e.g. \cite{Wilkinson_2009}, \cite{Wang-Ji_2018}). The spectral index $1.46^{+0.01}_{\rm -0.01}$ is found to be somewhat lower than the \nustar values. The corona temperature ($126^{+38}_{\rm -7}$ keV) and ionisation parameter ($\log \xi  =  3.69^{+0.03}_{\rm -0.22} \ \rm \log[erg.cm/s]$) are found to be close to the \nustar fit values. The lamp-post height ($4.1^{+0.7}_{\rm -1.8}R_{\rm g}$) is also found to be consistent with the \cite{Buisson_2019MNRAS.490.1350B} value. The \astrosat fit is also observed to prefer a closer inner disc. However, the iron abundance and the \textsc{xillverCp} ionisation parameter are pegged near the maximum allowed values. The maximality of the  \textsc{xillverCp} ionisation parameter is most likely due to the limited spectral capability of LAXPC (more specifically, in LAXPC20), as discussed in section \ref{sec:discussion}.
Additionally, the LAXPC residual shows a peak at $\sim 36$ keV, which is not present in CZTI residual, indicating an instrumental origin. The feature can be attributed to Xenon K-edge and can be taken care of with the inclusion of a \textsc{Gaussian} component \citep{Navin_2019}. We found that a \textsc{Gaussian} component of line energy $\sim$35.9 keV and width $\sigma \sim$0.8 keV sufficiently takes care of the feature. The resultant M1 model fit yields a $\chi$2/dof of 872.0/814. 

\begin{figure}
\centering
	\includegraphics[width=1.1\linewidth]{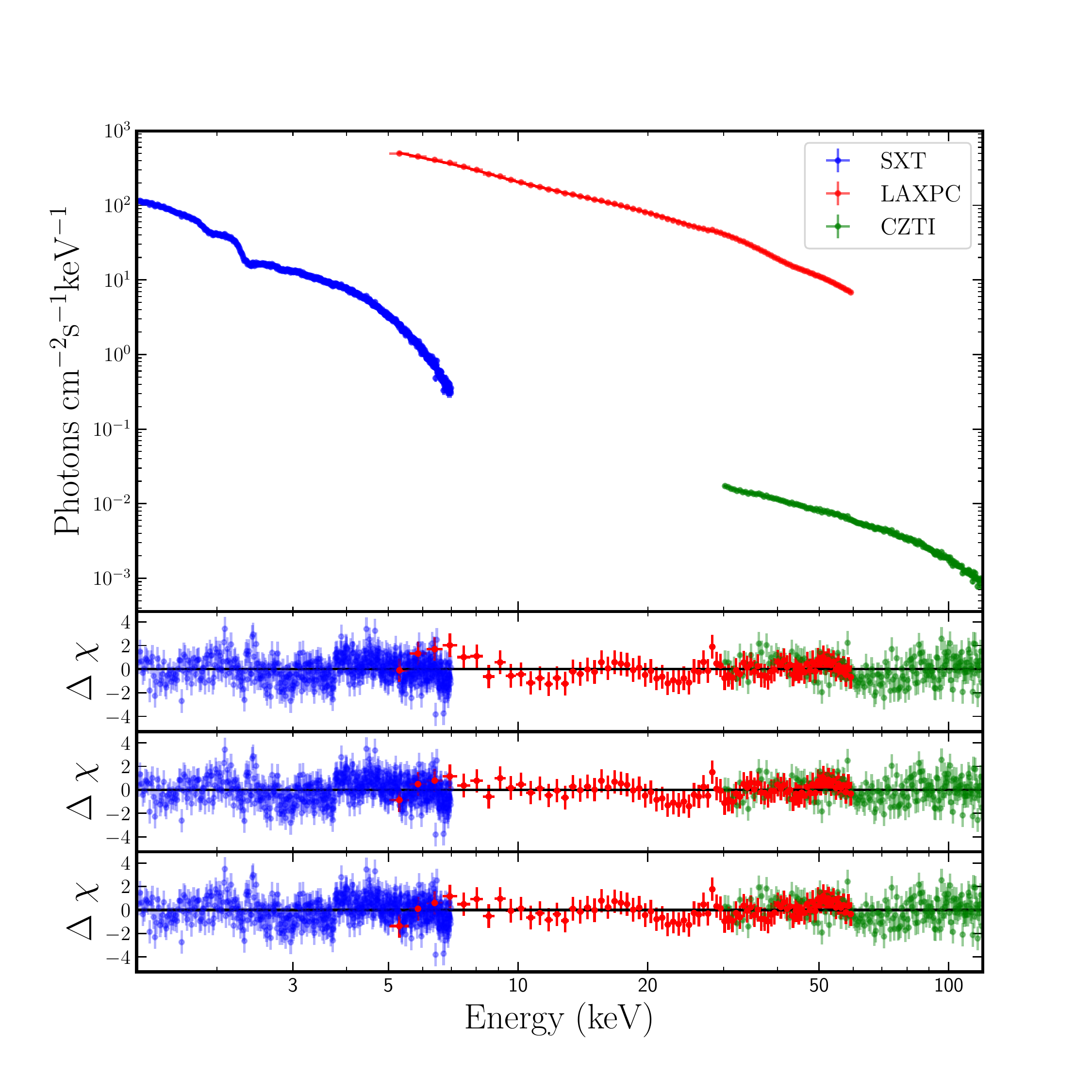}
  	\caption{Results of 1.3-120 keV \astrosat spectral fit of \maxi  (see sections~\ref{sec:w refl},~\ref{sec:discussion} for further details). Top panel: the \astrosat spectra. The blue, red and green points indicate SXT (1.3-7.0 keV), LAXPC (5-60 keV) and  CZTI (30-120 keV) data, respectively. The spectra are fitted with M2 model here. The difference in normalisation between SXT and the rest of the instruments can be attributed to small region of extraction we used to avoid pile-up (detailed in section~\ref{appendix:pile-up}). Second panel: Residuals from the M1 model fit. Third panel: Residuals from M2 model fit. Bottom panel: Residuals from M3 model fit.}
    \label{fig:astrosat}
\end{figure}

Similar to \nustar M2 model fit, we add a second \textsc{Nthcomp} component to the \astrosat M1 fit. The resultant M2 model fit yields a much better $\chi$2/dof of 856.8/8111. The best-fit \astrosat values are noted to be closer to the \nustar values for M2 model fit. The \textsc{diskbb} temperature remains almost unchanged. The lamp-post height ($3.18^{+0.38}_{\rm -0.47}R_{\rm g}$) is found to be consistent with \cite{ErinKara_2019Natur.565..198K}. Similar to \nustar, the fit with the M2 model results in a much lower value of the $kT_{\rm e}$ of the second Comptionisation ($18.4^{+3.6}_{\rm -3.2}$ keV), as opposed to the higher $kT_{\rm e}$ value ($149^{+79}_{\rm -33}$ keV) of the  primary Comptionisation component. The iron abundance and the \textsc{xillverCp} ionisation parameter are still found to be pegged at the upper limit. The spectral index of the second \textsc{Nthcomp} is found to be steeper ($1.57^{+0.01}_{\rm -0.05}$) than the intrinsic Comptonisation component ($1.41^{+0.03}_{\rm -0.04}$). The inner disc radius ($4.2^{+1.0}_{\rm -0.9}R_{\rm ISCO}$) also becomes comparable to the \nustar best-fit value. The best-fit parameter values for M1 and M2 fit are reported in the 3rd and 4th column in table ~\ref{tab:param}. The spectra and the residuals can be found in Fig. \ref{fig:astrosat}. The high cross-normalisation factors between SXT and LAXPC/CZTI can be attributed to the small extraction area for SXT (explained in greater detail in appendix~\ref{appendix:pile-up}). 
To further probe the effect of LAXPC on the significance of the second Comptonisation component, we fit the SXT+CZTI spectra between 1.3-5.0 keV and 40.0-120.0 keV with \textsc{TBabs$\times$(diskbb+Nthcomp)} and \textsc{TBabs$\times$(diskbb+Nthcomp+Nthcomp)} models. The former fit results in a $\chi$2/dof of 1165.2/741, while the latter gives a $\chi$2/dof of 824.8/738. This further establishes the importance of the second Comptonisation component.

Finally, for a closer parallel to the \astrosat data with the published \nustar{} fits, we adopt a similar model to \citet{Buisson_2019MNRAS.490.1350B}, involving a \textsc{diskbb} and two \textsc{relxilllpCp} components (denoted as \textsc{relxilllpCp(1)} for the reflection component with lower height, and \textsc{relxilllpCp(2)} for the higher corona). All the continuum parameters are allowed to vary freely. For simplicity, the inner disc radius is fixed at the ISCO. Thus, this model (M3: \textsc{TBabs$\times$(diskbb+relxilllpCp(1)+relxilllpCp(2))}) has two Comptonisation components at different heights, which are reflected from the inner disc. The resulting fit (model M3 in table \ref{tab:param}) can be compared to M2 model fit, with a $\chi^2$/dof of 861.4/818. The inclination angle is fixed at $30^{\circ}$, as allowing it to freely vary results in an unconstrained inclination value. The corona temperature shows a clear segregation, with a much higher temperature ($150^{+9}_{-7}$ keV) for the corona closer to the black hole and a much lower temperature ($22.9^{+7.4}_{-4.3}$ keV) for the corona component further away. The temperature of the colder corona is somewhat consistent wih \nustar and \astrosat M2 best-fit temperatures of the second Comptonisation component. The hotter corona (height of $2.8^{+0.8}_{-0.5} R_{\rm g}$) component is also found to have higher ionisation parameter than the colder corona (height of $5.5^{+2.3}_{-1.2} R_{\rm g}$) component. Note that, although the temperature structure is prominent, the segregation of lamp-post heights is not as pronounced as reported by  \citet{Buisson_2019MNRAS.490.1350B}.

\subsubsection{Estimation of mass of the Black Hole} \label{sec:mass}

The inner accretion disc radius in a black hole LMXB can be measured either from fitting the disc or from modelling the disc reflection. Assuming the inner radii deduced from the \textsc{diskbb} and \textsc{relxilllpCp} components to be the same, we can derive the mass of the black hole ($M_{\rm BH}$) from equation 3 in \cite{Wang_2018}. We use a correction factor of 1.2 \citep{Kubota_1998PASJ}, accounting for the spectral hardening \citep{Shimura_1995} and the fact that the disc temperature does not peak at the inner radius (\citet{Makishima_2000}, see also equation 2 and the corresponding discussion in \citet{Reynolds_2013}).

Using our best-fit values of $R_{\rm in}$ and \textsc{diskbb} normalisation from a fit with the \astrosat model M2, as well as considering a distance estimate from \cite{Gandhi_2019}, we get the mass of the black hole in \maxi to be $6.7-13.9 \ M_{\odot}$. This mass estimate from our spectral model is consistent with the estimates from dynamical mass measurement from \cite{Torres_massEstimate_2019ApJ...882L..21T} ($7-8 \ M_{\odot}$) and radio parallax measurement by \cite{Atri2020} ($9.2 \pm 1.3 \ M_{\odot}$). Additionally, this mass estimate from the hard state spectra is also consistent with the mass value derived from soft state spectra \citep{Fabian_2020}.

\section{Discussions} \label{sec:discussion}

In this paper, we report the results of broadband spectral analysis of 2018 hard state data of the transient black hole X-ray binary \maxi using all three pointing X-ray instruments (SXT, LAXPC and CZTI) on board \astrosat, as well as the nearest (both in time and in HID) available \nustar data. \nustar and \astrosat are the two most prominent currently active broadband X--ray satellites. While the \nustar data provides us with an opportunity to investigate the reflection features due to its superior energy resolution, the \astrosat data provides us a better handle over some other continuum parameters due to its much broader energy coverage between 1.3-120.0 keV. Thus, studying the source systematically in similar, contemporaneous states with the same models utilising two of the most prominent broadband instruments with such complementary capabilities, we derive conclusions that are more general and reliable. While the \nustar spectra are explored in great detail by \citet{Buisson_2019MNRAS.490.1350B}, they tied the two coronal temperature. We show, in section ~\ref{sec:w refl}, that this leads to a similar temperature even with our different model implementation. We further show that a much more general assumption of untied temperatures between the two coronal components leads to a better fit and segregation of the two temperatures. Similarly, the \astrosat spectrum has been presented in \citet{Mudambi_2020}. However, it is used to get a rough understanding of the spectrum to help in their main aim of spectro--timing study. That apart, their spectra did not include CZTI, a better understanding of SXT pile--up, and not a very good fit even with 3\% systematics (as opposed to the recommended 2\% we have used). The \textsc{diskbb} normalisation they derived, would imply to a black hole mass of 22--26 $M_{\odot}$, leading to inconsistent results. This motivated us to further explore the \astrosat data, with all three pointing X--ray instruments, in much greater detail.

The broadband spectra contains presence of soft excess, broad and narrow iron line complex and a Compton hump. The resultant spectrum is well fitted with the combination of a multi-colour disc black-body, a corona with a lamp-post geometry in the form of self-consistent, relativistic reflection model \textsc{relxilllpCp} (which contains an intrinsic thermal Comptonisation continuum), and a distant, unblurred reflection component in the form of \textsc{xillverCp}. The resultant model, M1, is detailed in section \ref{sec:w refl} and table \ref{tab:param}. From the best-fit \nustar spectra, we find the parameter values to be largely consistent with the previously reported values by \cite{Buisson_2019MNRAS.490.1350B} (epoch 3). The super-solar iron abundance of $4.4^{+1.3}_{\rm -0.5}$ (relative to solar values) is also found to be consistent with the previous results. Similar overabundance of iron has been reported in various X-ray binaries (e.g. \cite{Degenaar_2017},\cite{Garcia_2018}, \cite{Tomsick_2018}). The spectral index of the intrinsic \textsc{Nthcomp} is found to be $1.54-1.55$, typical of black hole X-ray binaries in the hard state. The distant reflection component is found to be ionised. The \textsc{diskbb} $T_{\rm in}$ is found to be $\sim 0.42-0.45$ keV, which is lower than the previously reported value by \cite{Buisson_2019MNRAS.490.1350B}. It is to be noted, however, that \nustar less reliable for the measurement of the \textsc{diskbb} temperature and normalisation. Additionally, the \nustar data fit with the M1 model reveals a higher corona temperature ($118^{+29}_{\rm -19}$ keV) than previously reported. For the spectral characterisation of \astrosat data, we jointly fit the SXT (corrected for pile-up and appropriately gain shifted), LAXPC (LAXPC20) and CZTI data, utilising the full broadband capability of \astrosat between 1.3-120.0 keV. The \astrosat best-fit with M1 model indicates a lower but better constrained \textsc{diskbb} $T_{\rm in}$ of $0.33^{+0.01}_{\rm -0.01}$ keV, a similar corona temperature and ionisation parameter, and a somewhat lower spectral index.

We also explore the possibility of a two-component corona through the addition of an external \textsc{Nthcomp} to our existing model. This implementation is similar in nature to the \textsc{relxilllpCp+relxilllpCp} scenario considered in \cite{Buisson_2019MNRAS.490.1350B}, but has some key differences. Both the ionisation parameter and the electron temperature of the two Comptonisation components (the intrinsic \textsc{Nthcomp} continuum of the \textsc{relxilllpCp} model and the added \textsc{Nthcomp}) are kept free and independent of each other. This gives us a more consistent and general picture, as the two components of the corona might have different physical properties.
The inclusion of this additional Comptonisation component (the M2 model) vastly improves the goodness of fit for both \nustar and \astrosat data. In case of \nustar, tying up the temperature of the two coronal components leads to a $kT_{\rm e}$ of $38^{+3}_{\rm -2}$ keV, consistent with \citet{Buisson_2019MNRAS.490.1350B}. Untying the two temperatures and letting both vary freely; leads to a much better $\chi^2$/dof than the previously reported models. The \textsc{diskbb} $T_{\rm in}$ is almost unaffected by the inclusion of this added component, and the other best-fit parameters are broadly consistent between \nustar and \astrosat. For both \nustar and \astrosat, the height of the lamp-post corona is found to be consistent with \cite{ErinKara_2019Natur.565..198K}. Letting both the corona temperatures free, leads to a segregation of temperatures for both \nustar and \astrosat. While the temperature of the primary Comptonisation component is found to be $115^{+38}_{\rm -29}$ keV and $149^{+79}_{\rm -33}$ keV, the $kT_{\rm e}$ of the second Comptonisation component is found to be $14.0^{+2.1}_{\rm -1.7}$ keV and $18.4^{+3.6}_{\rm -3.2}$ keV for \nustar and \astrosat data, respectively. This difference in corona temperatures can be interpreted as originating from different distances from the black hole. The high energy corona is much closer to the black hole, has a higher electron temperature and contributes to the broad iron line through blurred reflection; while the low energy corona is further away, has much lesser electron temperature and contributes to the narrow core of the iron line complex. Similar interpretation of inhomogeneous corona has been used for Cyg X-1 (\cite{Yamada_2013}, \cite{Basak_2017}). This idea is further reinforced by the implementation of our M3 model in table \ref{tab:param}, where for \astrosat data we used a model almost similar to \citet{Buisson_2019MNRAS.490.1350B}, with both the corona reflecting for a disc extending upto the ISCO and the continuum parameters of the two corona allowed to vary freely. This leads to a separation in both temperature and (to a certain extent) height, with the corona further away (\textsc{relxilllpCp(2)}) having lower ionisation and temperature than the corona closer to the back hole (\textsc{relxilllpCp(1)}). Note that this interpretation, though a little different, is not inconsistent with the contracting corona scenario \citep{ErinKara_2019Natur.565..198K}; it just assigns a more realistic temperature structure to the corona. This also shows that tying up the temperatures of the higher and lower temperature corona leads to an average temperature, similar to the one found by \citet{Buisson_2019MNRAS.490.1350B}.

A few points are to be noted about the fits described in our work. First of all, by setting the \textsc{relxilllpCp} and \textsc{xillverCp} refl\_frac$<0$ (thereby including only the reflected flux from the respective components), we attempt to avoid the possibility of the second reflection component adding significantly to the continuum (and hence mimicking the spectra for a second corona temperature), rather than only fitting the reflection features (iron line, Compton hump etc.). The similarity in the best-fit parameters for \nustar and \astrosat, and the likelihood of the second Comptonisation with a similar temperature for both the instruments provide a further credence to our claim of a multi-temperature corona. That apart, the M3 model fit of \astrosat data provides further support that the two different temperatures can be attributed to different corona heights. Nevertheless, the alternate scenario (only one temperature and second reflection component adding to the continuum) cannot be entirely ruled out.

As \textsc{xillverCp} corresponds to a corona situated farther away, this component should be less ionised than the  \textsc{relxilllpCp} component. This is what we find for \nustar fits (see Table~\ref{tab:param}). However, the \textsc{xillverCp} ionisation parameter, for \astrosat fits, is pegged to a maximal value, and is higher than the best-fit value of the \textsc{relxilllpCp} ionization parameter (see Table~\ref{tab:param}). This higher value could be because \astrosat (LAXPC20, in particular) cannot adequately describe the narrow features of \textsc{xillverCp} (see Fig.~\ref{fig:eemo}) due to a limited spectral capability.
However, the \textsc{xillverCp} component is necessary even for \astrosat fits. We conclude this, because, while the $\chi^2$/dof is $856.8/811$ for the model M2, it is $885.1/814$ if \textsc{xillverCp} is excluded from M2. Note that, if we force a lower value of the \textsc{xillverCp} ionisation parameter, by tying it to the \textsc{relxilllpCp} ionisation parameter for the M2 model of \astrosat fitting, the contribution of the \textsc{xillverCp} component becomes negligible, and the fit is much worse with $\chi^2{\rm /dof} = 886.4/813$. Nevertheless, the best-fit parameter values (e.g., \textsc{diskbb} $T_{\rm in} = 0.29_{-0.01}^{+0.01}$ keV) for this modified M2 model are overall consistent with those for the M2 model for \astrosat fitting. Moreover, the fits of both \astrosat and \nustar spectra with the M2 model give a similar conclusion, e.g., a temperature structure of the corona. These give confidence to our results, and we list the best-fit parameter values for the \astrosat M2 model fit in Table~\ref{tab:param}, although \astrosat cannot adequately model the \textsc{xillverCp} component.

The $R_{\rm in}$ for both \nustar and \astrosat are found to be of similar values ($5.4^{+1.5}_{\rm -1.5} R_{\rm g}$ for \nustar, and $4.2^{+1.0}_{\rm -0.9} R_{\rm g}$ for \astrosat for M2 model fit; see Table~\ref{tab:param}). For a maximally spinning black hole, this would place the inner edge of the accretion disc at $2.7-5.6 \ R_{\rm ISCO}$ ($R_{\rm ISCO}$: radius of the innermost stable circular orbit). However, this does not necessarily imply a truncated disc, as $R_{\rm in}$ and $a$ are degenerate and we fix the dimensionless spin parameter ($a$) to 0.998, only to simplify a stable fitting process (section~\ref{sec:w refl}). Note that $R_{\rm ISCO}$ is determined by the black hole spin, and it monotonically increases from $1.24R_{\rm g}$ for an extremely prograde spinning black hole to $9.0R_{\rm g}$ for an extremely retrograde spinning black hole. 
The alternate scenario of a low black hole spin with a disc extending all the way down to ISCO, is equally likely, and the spectral fitting alone may be insufficient to distinguish between different $a$-values. In fact, the almost unchanging $R_{\rm in}$ from \citet{ErinKara_2019Natur.565..198K} and the consistency of best-fit $R_{\rm in}$ throughout the 8 epoch in \citet{Buisson_2019MNRAS.490.1350B} imply that $R_{\rm in}$ does not vary much throughout the outburst. Furthermore, the soft state spectral fitting by \citet{Fabian_2020} supports a spin value between -0.5 and +0.5. Both these evidences indicate that \maxi{} may contain a low spin black hole, with the disc extending to ISCO. To test if our best-fit parameter values allow this possibility, we fix the $R_{\rm in}$ to the ISCO radius, make the spin parameter $a$ free, fix the inclination at $30^{\circ}$, and freeze all the other parameters (except the normalisations) to their best-fit values in the M2 model fit of \nustar data.
This results in an acceptable fit, with a $\chi^2$ of 672.0 for 678 degrees of freedom. The corresponding best-fit spin parameter is $0.48^{+0.25}_{-0.26}$, which is consistent with \citet{Fabian_2020}.

Finally, assuming the same inner disc radius between the \textsc{diskbb} and \textsc{relxilllpCp}, we calculate the mass of the black hole to be $6.7-13.9 \ M_{\odot}$, consistent with previously reported values. This is the first time the mass of this black hole has been calculated from the hard state spectra in such a way. It is to be noted that most studies deriving the color-correction factor have involved BHXBs predominantly in soft state (see \citet{Reynolds_2013,Merloni_2000} for exceptions), and a higher value might be warranted in hard states \citep{Reynolds_2013,Dunn_2011,Davis_2019}. The consistency of the black hole mass estimate through this method with the other values, however, can provide a justification towards a disc origin of the thermal emission (as opposed to a reflection origin). This is further corroborated by \citet{Fabian_2020}.

From our best-fit model (M2), we find the unabsorbed 0.1-500 keV flux to be $1.64-1.69 \times 10^{-7} \ \rm{erg/cm^{2}/s}$ for the \nustar data and $1.62-1.64 \times 10^{-7} \ \rm{erg/cm^{2}/s}$ for the \astrosat data. This implies a 0.1-500 keV luminosity of $L_{\rm 0.1-500 \ \rm{keV}} \sim 2.32-2.42 \times 10^{38} \ \rm{erg/s}$. The implies that the black hole in \maxi is accreting at $13-29 \% \ L_{ \rm Edd}$.

To summarise, through systematic investigation of contemporaneous \astrosat and \nustar data in similar states and with similar models, we have characterised the broadband X-ray spectra of the transient black hole X-ray binary \maxi in the hard state (March 2018) during its 2018 outburst. The \nustar best-fit parameters are found to be largely consistent with the values reported in the literature. We also fully utilise the broadband capability of all the pointing X-ray instruments on board \astrosat, through consistent spectral analysis in 1.3-120.0 keV energy range. Though there are some quantitative differences of the best-fit parameters between \nustar and \astrosat, the broad conclusions are consistent with each other. The inclusion of \astrosat data complements the \nustar data, as \nustar spectral fit gives us a better estimate of the reflection parameters, while \astrosat data provides superior estimates of disc temperature, normalisation and the high energy rollover of the corona. We utilise this potential to also investigate the possibility of an inhomogeneous corona through the implementation of a double Comptonisation model, which leads to better goodness of fit for both \nustar and \astrosat data. The consistency of the low energy Comptonisation component for independent observations with different instruments in a similar state (with \astrosat covering a broader energy range) establishes the significance of our results.

\begin{table*}
\caption{Parameters of fits to \maxi\ spectra in the hard state observations  with \astrosat and \nustar. The models M1, M2 and M3 are detailed in section~\ref{sec:w refl}. Errors represent 90\% confidence intervals. $^f$ denotes that the corresponding parameter is frozen. $^p$ or $_p$ denotes that the parameter is pegged at the upper/lower limit value.}
\label{tab:param}
\begin{tabular}{lllcccccccc}
\hline
Spectral Component & Parameter & \multicolumn{2}{c}{\nustar}  & \multicolumn{3}{c}{\astrosat} \\ 
& & M1 & M2 & M1 & M2 & M3 \\
\hline

\verb'diskbb' & $kT_\mathrm{in}$ (keV) & $0.44_{-0.07}^{+0.08}$ (FPMA) & $0.45_{-0.05}^{+0.06}$ (FPMA) & $0.33_{-0.01}^{+0.01}$ & $0.31_{-0.01}^{+0.01}$ & $0.35_{-0.01}^{+0.01}$ \\
[1ex]
& & $0.42_{-0.08}^{+0.09}$ (FPMB) & $0.43_{-0.05}^{+0.05}$ (FPMB) &  &  & \\
[1ex]   
    & norm $(\times 10^4)$ & $2.25_{-0.99}^{+4.37}$ (FPMA) & $1.74_{-0.44}^{+7.54}$ (FPMA) & $1.39_{-0.26}^{+0.36}$ & $1.62_{-0.35}^{+0.49}$ &  $1.21_{-0.24}^{+0.34}$\\
    [1ex]
    & & $2.02_{-1.13}^{+21.4}$ (FPMB) & $1.67_{-0.42}^{+1.26}$ (FPMB) &  & & \\

&&&&&&&\\

\verb'Nthcomp(1)' & $\Gamma$ & $1.54_{-0.01}^{+0.01}$ (FPMA) & $1.49_{-0.04}^{+0.03}$ (FPMA) & $1.46_{-0.01}^{+0.01}$ & $1.41_{-0.04}^{+0.03}$ & ... \\
[1ex]
& & $1.54_{-0.01}^{+0.01}$ (FPMB) & $1.49_{-0.08}^{+0.02}$ (FPMB) &  &  & \\
[1ex]
    &$kT_\mathrm{e}$ (keV) & $118_{-19}^{+29}$ & $115_{-29}^{+38}$ & $126_{-7}^{+38}$ & $149_{-33}^{+79}$ & ... \\
&&&&&&&\\

\verb'relxilllpCp(1)' & $\Gamma$ & ... & ... & ... & ... & $1.38_{-0.01}^{+0.01}$\\
\verb'(lower reflection)' & & & & & &  \\
[1ex]
    &$kT_\mathrm{e}$ (keV) & ... & ... & ... & ... & $150_{-7}^{+9}$ \\
    [1ex]
    & \textit{h} ($R_\mathrm{g}$) & $<2.8$ & $7.41_{-1.93}^{+1.94}$ & $4.1_{-1.8}^{+0.7}$ & $3.2_{-0.5}^{+0.4}$ & $2.8_{-0.5}^{+0.8}$ \\
[1ex]
    & \textit{$\theta$} ($^\circ$) & $29_{-8}^{+2}$ & $25_{-2}^{+8}$ & $27_{-10}^{+8}$ & $35_{-9}^{+7}$ & $30^{f}$\\
[1ex]
    & $R_\mathrm{in}$ ($R_\mathrm{g}$) & $6.9_{-1.0}^{+0.9}$ & $5.4_{-1.5}^{+1.5}$ & $2.5_{-1.0}^{+1.2}$ & $4.2_{-0.9}^{+1.0}$ & $1.2^f$ \\
   [1ex]
    & $\log{\xi}$ (log[erg cm/s]) & $3.9_{-0.1}^{+0.2}$  & $3.38_{-0.03}^{+0.04}$ & $3.69_{-0.22}^{+0.03}$ & $3.68_{-0.13}^{+0.03}$ & $4.21_{-0.01}^{+0.01}$ \\
    [1ex]
    & $A_{\rm Fe}$ ($A_{\rm Fe,\odot}$) & $4.4_{-0.5}^{+1.3}$  & $5.0_{-0.2}^{+0.3}$ & $10.0^p$ & $10.0^p$ &  $10.0^p$ \\
    [1ex]
    & $\mathcal{R}$  & $0.54$ & $1.21$ & $1.93$ & $1.30$ & $2.42$ \\
    [1ex]
    & norm & $1.96$ (FPMA) & $0.09$ (FPMA) & $0.05$ & $0.12$ & $0.16$ \\
    &  & $1.70$ (FPMB) & $0.08$ (FPMB) &  & & \\
&&&&&&&\\

\verb'xillverCp'  & $\log{\xi}$ (log[erg cm/s]) & $2.5_{-0.1}^{+0.2}$ & $2.4_{-0.2}^{+0.1}$  & $4.70^p$ & $4.70^p$ & ... \\
[1ex]
                & norm $(\times 10^{-2})$ & $1.4$ & $0.4$ & $2.36$ & $1.16$ & ... \\

&&&&&&&\\

\verb'Nthcomp(2)' & $\Gamma$ & ... & $1.66_{-0.02}^{+0.02}$ & ... & $1.57_{-0.05}^{+0.01}$ & ... \\
[1ex]
   &$kT_\mathrm{e}$ (keV) & ... & $14.0_{-1.7}^{+2.1}$ & ... & $18.4_{-3.2}^{+3.6}$ & ... \\
   [1ex]
   & norm & ... & $2.17$ & ... & $0.40$ & ... \\

&&&&&&&\\

\verb'relxilllpCp(2)' & \textit{h} ($R_\mathrm{g}$) & ... & ... & ... & ... & $5.5_{-1.2}^{+2.3}$ \\
[1ex]
\verb'(upper reflection)'   & $\Gamma$ & ... & ... & ... & ... & $1.44_{-0.01}^{+0.01}$\\
[1ex]
&$kT_\mathrm{e}$ (keV) & ... & ... & ... & ... & $22.9_{-4.3}^{+7.4}$ \\
  [1ex]
   & $\log{\xi}$ (log[erg cm/s]) & ... & ... & ... & ... & $3.62_{-0.07}^{+0.09}$ \\
[1ex]
    & $\mathcal{R}$  & ... & ... & ... & ... & $1.92$ \\
        [1ex]
    & norm & ... & ... & ... & ... & $0.01$ \\
    
\hline\\

$\chi^2/{\rm d.o.f.}$  & & $715.1/670$ & $666.1/667$ & $872.0/814$ & $856.8/811$ & $861.4/818$ \\

\hline\\

    & $C_\mathrm{LAXPC}$ & ... & ... & $1.95$ & $1.95$ & $1.93$ \\
[1ex]
    & $C_\mathrm{CZTI}$ & ... & ... & $1.94$ & $1.94$ & $1.93$ \\
[1ex]
    & $C_\mathrm{FPMB}$ & $0.96$ & $0.96$ & ... & ... & ... \\

\hline\\

Unabsorbed flux & 3.0--70.0 keV & 8.9 & 8.9 & 7.8 & 7.8 & 7.8 \\

\hline\\

\end{tabular}
 \begin{flushleft}
 
 \textbf{Note:}
 $T_\mathrm{in}$: Temperature of the inner disc; norm: Normalisation of the corresponding spectral parameter; $\Gamma$: Asymptotic power-law photon index; $T_\mathrm{e}$: Electron temperature of the corona, determining the high energy rollover; \textit{h}: Height of the comptonising lamp-post corona above the black hole; $\theta$: Inclination of the inner disc; $R_\mathrm{in}$: Inner disc radius (in units of $R_\mathrm{\rm g}$); $\xi$: Ionisation parameter of the accretion disc, defined as $\xi=L/nR^{2}$, with \textit{L}, \textit{n}, \textit{R} being the ionising luminosity, gas density and the distance to the ionised source, respectively; $A_{\rm Fe}$: Iron abundance, in the units of solar abundance; $\mathcal{R}$: Reflection fraction; $C_\mathrm{LAXPC}$: the flux normalisation constant for LAXPC (determined by multiplicative \texttt{`constant'} parameter in the spectral models) is estimated with respect to the SXT flux. Similarly, we have denoted $C_\mathrm{CZTI}$ for CZTI flux normalisation constant with respect to the SXT flux, and $C_\mathrm{FPMB}$ for FPMB flux normalisation constant with respect to the FPMA flux.

\end{flushleft}
\end{table*}

\section*{Acknowledgements}
We thank the referee for constructive comments, which improved the paper.
We also acknowledge the supports from Indian Space Research Organisation (ISRO) for mission operations and distributions of the data through ISSDC. This work has used the data from the Soft X-ray Telescope (SXT) developed at TIFR, Mumbai, and the SXT Payload Operations Center (POC) at TIFR is thanked for verifying and releasing the data via the ISSDC data archive and providing the necessary software tools. The LAXPC POC, TIFR, Mumbai is also acknowledged for providing us important inputs and necessary tools for data analysis. This research has made use of the $MAXI$ data provided by RIKEN, JAXA and the $MAXI$ team. This research has also made use of the \nustar Data Analysis Software (NuSTARDAS), jointly developed by the ASI Science Data Center (ASDC, Italy) and the California Institute of Technology (USA). The authors also thank Dr. Sunil Chandra for valuable insights and comments regarding SXT pile-up correction.

\section*{Data Availability}

The observational data used in this paper are publicly available at NASA's High Energy Astrophysics Science Archive Research Center (HEASARC; \url{https://heasarc.gsfc.nasa.gov/}) and ISRO's Science Data Archive for AstroSat Mission (\url{https://astrobrowse.issdc.gov.in/astro_archive/archive/Home.jsp}), and references are mentioned. Any additional information will be available upon reasonable request.




\bibliographystyle{mnras}
\bibliography{maxi_j1820} 



\appendix

\section{Pile-up check}\label{appendix:pile-up}

The SXT PC mode data for \maxi, taken at a time when the source flux is >1 Crab, is found to be highly piled up. In order to investigate and rectify the pile-up effect, we fit the 1.3-5.0 keV SXT data with a \textsc{TBabs $\times$ (diskbb+Nthcomp)} model, avoiding the Fe K-$\alpha$ complex. The $N_H$ is fixed to $1.5 \times 10^{21} \ \rm cm^{-2}$, as used in the rest of the work, and the \textsc{Nthcomp} seed temperature is set at the $T_{\rm in}$ of the \textsc{diskbb} component. This model is applied to different groups of SXT data, each group differing from previous one by the greater amount of area of the central bright region excluded from the source PSF on the SXT CCD., with the outer edge of the selected annular regions fixed at $15^{'}$. Similar to the procedure in \citet{Romano_2006}, it is noted that the spectra become steeper as more area is excluded until it reaches a stable value. The radius from the centroid of the PSF, where the spectral index reached a stable value, is chosen to be the inner radius of the annular region. This annular region from $6^{'}$ to $15^{'}$, is used as the source region throughout the rest of the work. 

\begin{figure}
\centering
    \includegraphics[width=\linewidth]{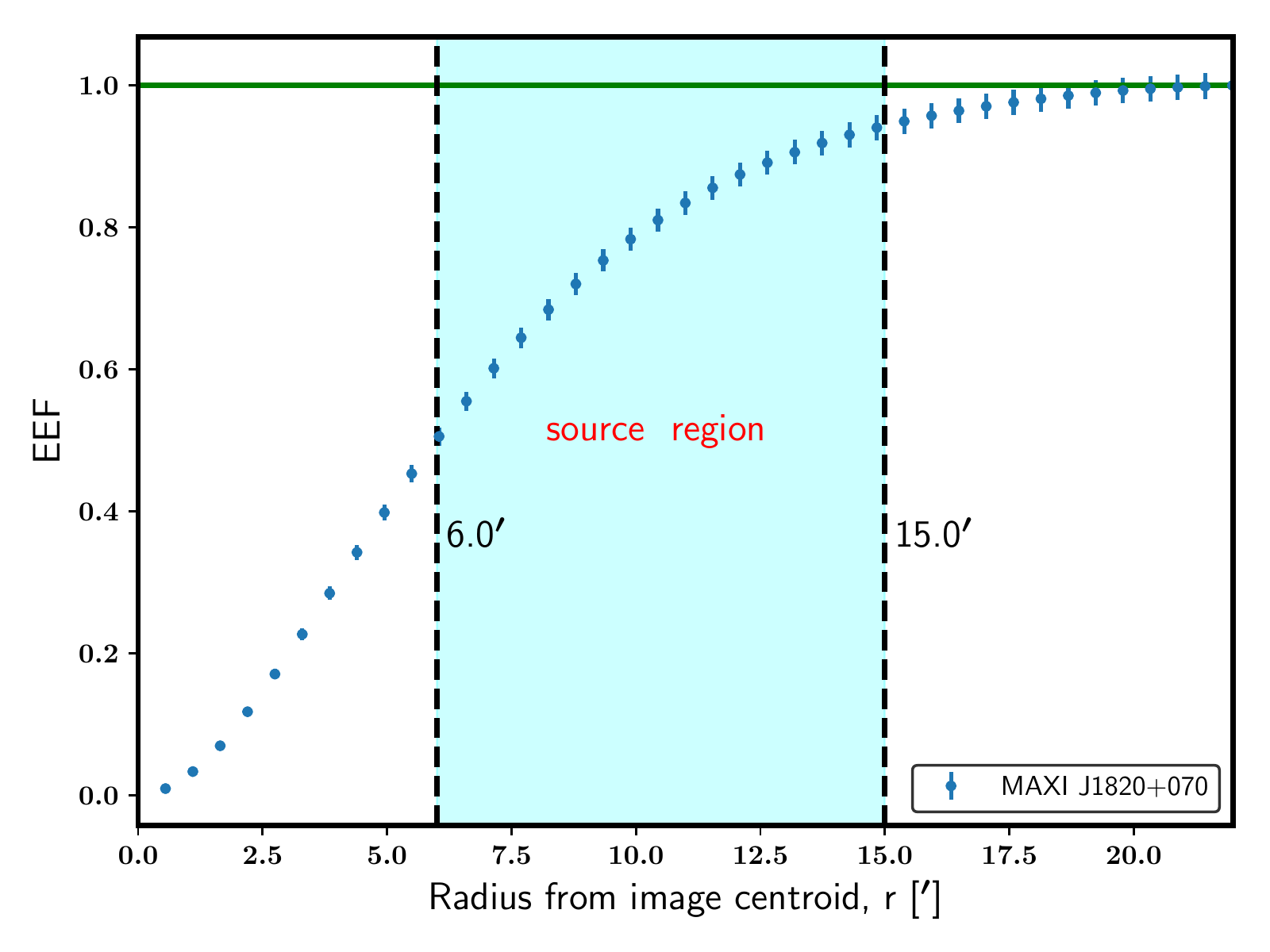}
      \caption{Encircled Energy Fraction (EEF) of the SXT data of \maxi, as a function of distance from the centroid of the PSF. The error bars have been multiplied by a factor of 50 for better representation. The shaded region between 6$^{'}$ and 15$^{'}$ is the chosen source region for our study.}
    \label{eef}
\end{figure}

Fig.~\ref{eef} shows a plot of the Encircled Energy Fraction (EEF) for SXT data of \maxi. From this plot, we can see that about 44\% of photons are encircled within the chosen source region, while a circular region with a radius of $15^{'}$ from the centroid of the PSF should contain about 94\% of the photons. Thus, the SXT flux is underestimated by a factor of $\sim 2$. This explains the large difference of normalisation factor between the SXT and the other two instruments.

\begin{figure}
\centering
    \includegraphics[width=\linewidth]{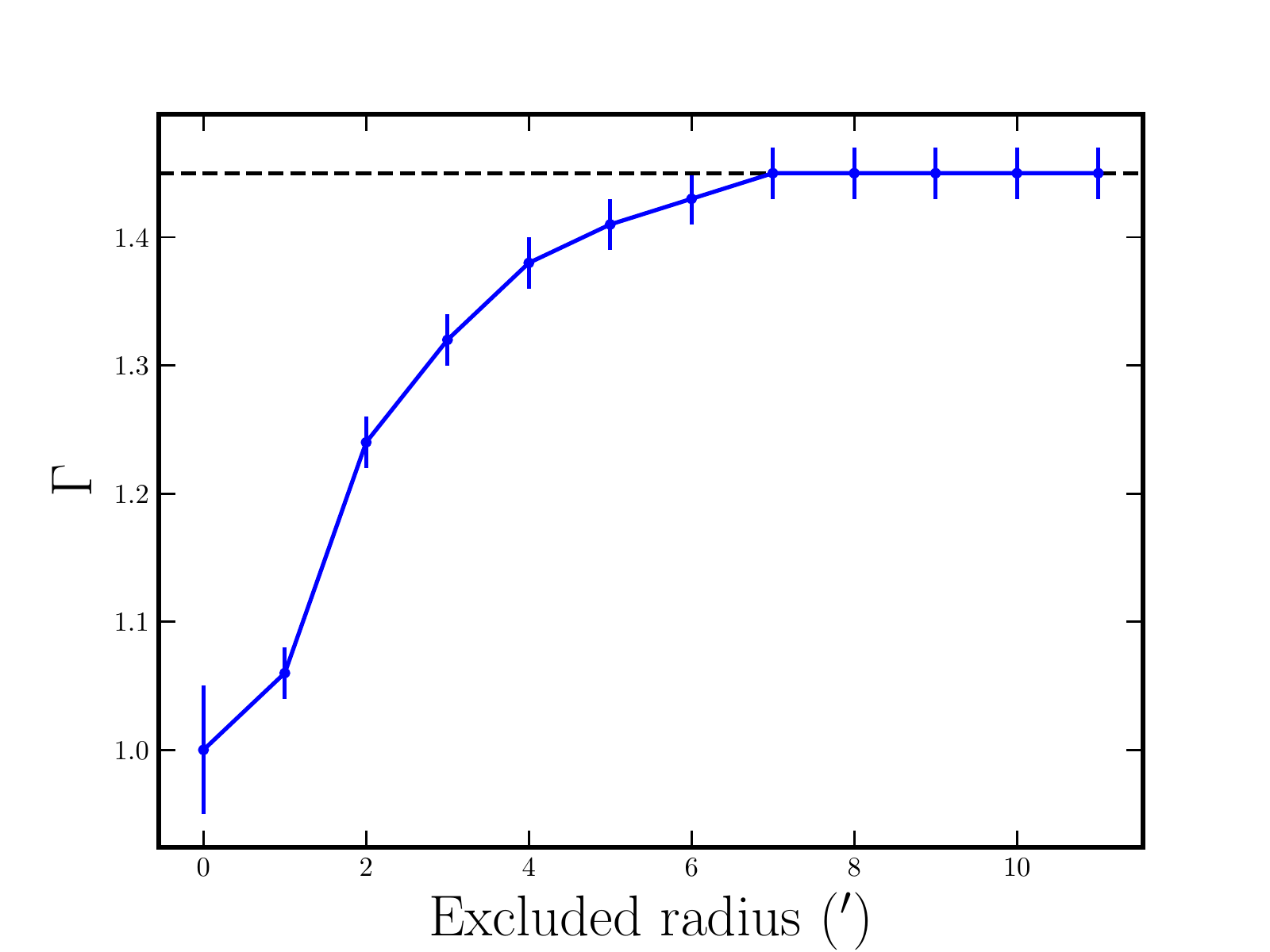}
      \caption{Plot to check the effect of pile-up, following \citet{Romano_2006}. We use different annular regions in the CCD image, with outer radius fixed at 15$^{'}$ from the centroid of the PSF and inner radius progressively excluded. Each spectrum is then fitted with a simple model described in Appendix \ref{appendix:pile-up}. The data points show the spectral index ($\Gamma$) as a function of the radius of the inner excluded region. We can observe that the spectral index increases and reaches a stable value roughly the same as the M1/M2 fit parameters. We have, therefore, chosen 6$^{'}$ as the inner radius of the selected source region for the spectral study in section~\ref{sec:w refl}.}
    \label{delc}
\end{figure}

\bsp	
\label{lastpage}
\end{document}